# Stark Effects of Rydberg Excitons in a Monolayer WSe$_2$ P-N Junction

Zhen Lian[1,2#], Yun-Mei Li[3#], Li Yan[2#], Lei Ma[2], Dongxue Chen[2], Takashi Taniguchi[4], Kenji Watanabe[5], Chuanwei Zhang[6,7], Su-Fei Shi[1,2*]

1. Department of Physics, Carnegie Mellon University, Pittsburgh, PA 15213, USA
2. Department of Chemical and Biological Engineering, Rensselaer Polytechnic Institute, Troy, NY 12180, USA
3. Department of Physics, School of Physical Science and Technology, Xiamen University, Xiamen, 361005, China
4. International Center for Materials Nanoarchitectonics, National Institute for Materials Science, 1-1 Namiki, Tsukuba 305-0044, Japan
5. Research Center for Functional Materials, National Institute for Materials Science, 1-1 Namiki, Tsukuba 305-0044, Japan
6. Department of Physics, University of Texas at Dallas, Richardson, TX 75080, USA
7. Department of Physics, Washington University in St Louis, St. Louis, MO 63105, USA

[#] These authors contributed equally to this work
[*] Corresponding authors: sufeis@andrew.cmu.edu

## Abstract

**The enhanced Coulomb interaction in two-dimensional (2D) semiconductors leads to the tightly bound electron-hole pairs known as excitons. The large binding energy of excitons enables the formation of Rydberg excitons with high principal quantum numbers (n), analogous to Rydberg atoms. Rydberg excitons possess strong interactions among themselves, as well as sensitive responses to external stimuli. Here, we probe Rydberg exciton resonances through photocurrent spectroscopy in a monolayer WSe$_2$ p-n junction formed by a split-gate geometry. We show that an external in-plane electric field not only induces a large Stark shift of Rydberg excitons up to quantum principal number n=3 but also mixes different orbitals and brightens otherwise dark states such as 3p and 3d. Our study provides an exciting platform for engineering Rydberg excitons for new quantum states and quantum sensing.**

**Introduction**

The reduced screening in 2D enhances the Coulomb interaction and gives rise to tightly bound electron-hole pairs, known as excitons, in monolayer transition metal dichalcogenides (TMDCs)[1–5]. Excitons in TMDCs exhibit enhanced light-matter interaction and play an important role in the optical and optoelectronic properties of monolayer TMDCs[6]. Excitons are tunable by external stimuli such as electrostatic doping[7,8], electric field[9,10], and magnetic field[11–13], making them promising in applications ranging from solar energy harvesting to quantum information science[6,14–16]. Enhancing exciton-exciton interactions and the response of excitons to external stimuli is critical for realizing those applications.

Rydberg exciton is a natural choice as it can be sensitively controlled through an external electric field owing to the large dipole moment. Similar to the energy states in a hydrogen atom, the energy states of an exciton can be described by a series of discrete states known as the excitonic Rydberg series, distinguished by their principal quantum number n and angular quantum number l[17,18]. Rydberg excitons refer to the excitons with high-order quantum principal numbers (n>1), analogous to Rydberg atoms. Previous works have demonstrated that the ground state, aka the 1s exciton, exhibits a quadratic Stark shift and tunable dissociation rate under an in-plane electric field[9]. We expect a much-enhanced Stark shift from Rydberg excitons due to their more extended wavefunction compared with 1s exciton.

In this work, we measure the photocurrent spectra[9,11,19–21] of a monolayer $WSe_2$ p-n junction formed by a split gate geometry, with the electric field tunable via the combination of the two gate voltages. We reveal the excitonic Stark effect of Rydberg excitons with n up to 3. Remarkably, the electric field results in the mixing of the optically bright s states and the otherwise dark p and d states, resulting in new bright exciton states with hybridized orbitals that are highly tunable by the electric field. The excited states exhibit energy shift and splitting as large as 96 meV (2s and 2p exciton under the electric field of 15 mV/nm), which are orders of magnitude larger than the shift of the ground state (1s) exciton (~ 2 meV) due to their larger radii. Our results show that the 2D Rydberg excitons are highly tunable by the in-plane electric field.

**Results and discussions**

**Photocurrent response from $WSe_2$ p-n junction**

We fabricated the monolayer $WSe_2$ device in a split-gate configuration, as shown in Fig. 1a. A typical device is composed of a monolayer $WSe_2$ connected by two few-layer graphene (FLG) as contact electrodes, sandwiched by the top and bottom h-BN layers. The final device is placed on two 3/22nm Ti/Au gate electrodes laterally separated by a gap of ~240 nm, which serve as two back gate electrodes of the device (gate 1 and gate 2). The split gate configuration allows the independent control of the doping levels of the two areas of the monolayer $WSe_2$ above gate electrodes 1 and 2 (region 1 and region 2) by applying independent gate voltages ($V_{g1}$ and $V_{g2}$) to the two gate electrodes. We then focus a laser beam with the photon energy of 1.96 eV (filtered output from a white laser) on the spot between regions 1 and 2, with the beam spot size about 4 µm, and collect the photocurrent response. A color plot of the photocurrent response as a function of $V_{g1}$ and $V_{g2}$ measured with 100 µW laser excitation from a typical device D1 at 77 K is shown in Fig. 1c. Clearly, the photocurrent response can be separated into four distinct regions (quadrants), separated by the lines corresponding to $V_{g1} = V_{g1}^0$ and $V_{g2} = V_{g2}^0$ in Fig. 1c. In the PN quadrant of Fig. 1c ($V_{g1} < V_{g1}^0$ and $V_{g2} > V_{g2}^0$), region 1 is p-doped while region 2 is n-doped. The device exhibits a positive photocurrent response due to the photovoltaic effect arising from the built-in field of the p-n junction (schematically shown in Fig. 1b). On the other hand, the NP quadrant ($V_{g1} > V_{g1}^0$ and $V_{g2} < V_{g2}^0$) corresponds to the n-p configuration, and the photocurrent switches the sign to be negative since the p-n junction reverses the direction of the built-in electric field. The photocurrent responses from the other two quadrants can be explained by the p-p and n-n configurations: The built-in field is small in these cases, and so is the photocurrent magnitude.

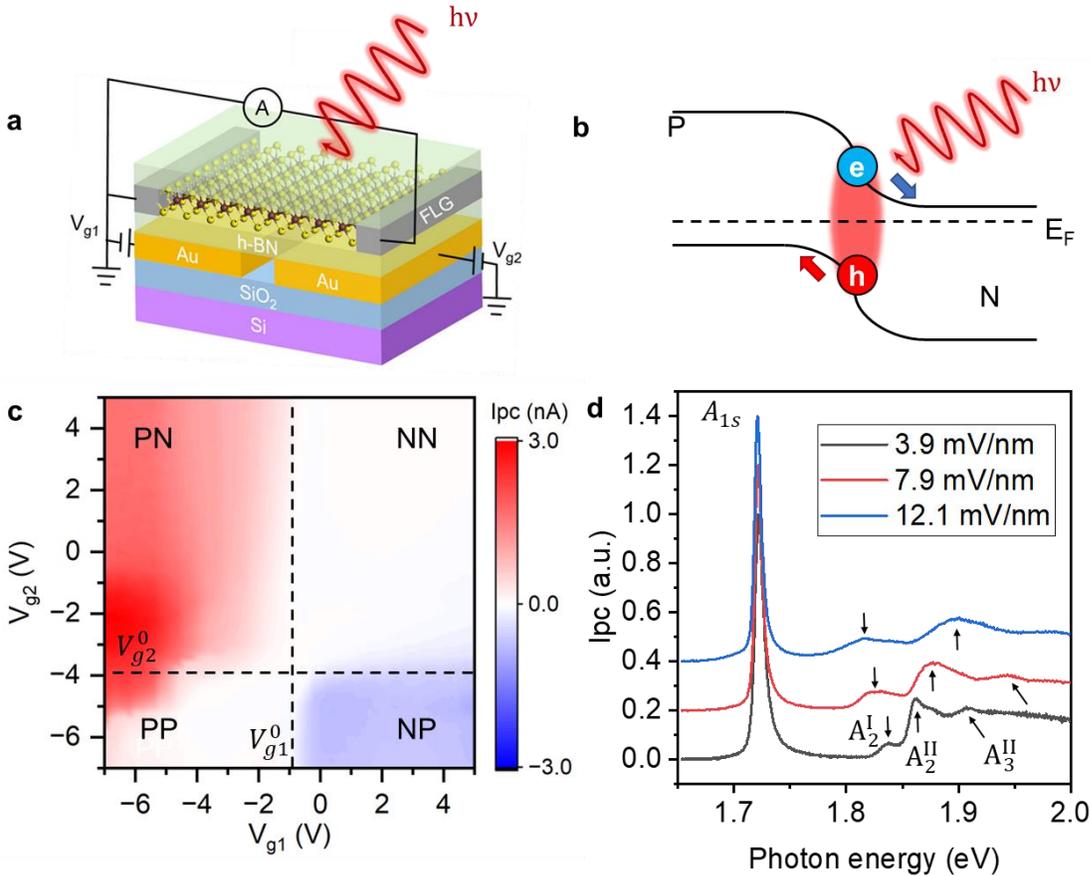

**Figure 1. Photocurrent spectra from a monolayer WSe₂ p-n junction.** (a) A schematic of the WSe₂ split-gate device. (b) A schematic of the band diagram and the photovoltaic effect of the WSe₂ p-n junction. (c) The photocurrent response from device D1 as a function of both V$_{g1}$ and V$_{g2}$. The photocurrent was measured using a pulsed laser excitation centered at 1.96 eV with a power of 100 uW. (d) Photocurrent spectra measured from device D1 under different electric fields, with an offset of 0.2 for better illustration. The magnitude of the photocurrent in (d) is normalized by the magnitude of the 1s exciton. All data shown in Fig. 1 were measured at 77 K.

We then focus on the PN quadrant in Fig. 1c and discuss the electric field dependence of the photocurrent spectra. We applied opposite voltages to gate 1 and gate 2 ($V_{g2} = -V_{g1} = V_g$) to apply the in-plane electric field. As we increase the value of $V_g$, we expect the width of the depletion region of the PN junction to decrease due to the increased doping level in regions 1 and 2, resulting in an enhanced built-in electric field strength. Using an analytical approach[9,22,23], we calculate the lateral electric field strength $F$ as a function of $V_g$, with details shown in the supplementary information (SI) section 2. In Fig. 1d we show the photocurrent spectra measured at three different electric field strengths, 3.9 mV/nm, 7.9 mV/nm, and 12.1 mV/nm. Each peak on the photocurrent spectra represents the absorption resonance of an excitonic state. The most prominent feature on the spectra is a strong absorption peak at 1.721 eV, which corresponds to the 1s state of WSe$_2$ A-exciton and is denoted as A$_{1s}$. It is worth noting that we do not observe any trion peaks, which are typically 20-30 meV below the A-exciton peak[24,25], suggesting the photocurrent response arises selectively from the depletion region of the junction. At the energies higher than the resonance of A$_{1s}$, we observe three weaker excitonic peaks at $F = 3.9 \, mV/nm$, which are denoted as $A_2^I$, $A_2^{II}$ and $A_3^{II}$ and are located at 1.838 eV, 1.863 eV, and 1.908 eV, respectively. $A_2^I$ and $A_2^{II}$ exhibit considerable energy shifts in opposite directions under an increasing electric field $F$. The redshift of $A_2^I$ and the blueshift of $A_2^{II}$ are 8 meV and 14 meV, respectively, at $F = 7.9 \, mV/nm$ and further increase to 21 meV and 38 meV, respectively, at $F = 12.1 \, mV/nm$. The blueshift of $A_3^{II}$ is even larger in magnitude, reaching 70 meV at $F = 12.1 \, mV/nm$.

**Stark effects of excitons**

To quantitatively understand the shift of each excitonic peak, we plot the detailed electric-field-dependent photocurrent spectra in Figs. 2a, b and c. Characteristic linecuts from the color plots are also shown in Fig. 2d. It is evident that the peak position of A$_{1s}$ (black dots in Fig. 2a) is a quadratic function of $F$, which can be expressed by $E_{1s} = E_{1s}^0 - \alpha F^2$, where $E_{1s}^0$ is the energy of A$_{1s}$ at zero electric field. From the fitting, we determine $\alpha = 8.0 \pm 0.2 \, eV/(V/nm)^2$, consistent with what has been extracted from the photocurrent spectra of A exciton as a function of electric field ($10 \pm 2 \, eV/(V/nm)^2$)[9]. In the electric field range between 4 and 15 mV/nm, the energy shifts of $A_2^I$ and $A_2^{II}$ are approximately linear functions of $F$, given by $E_n = E_n^0 + \beta F$ (Fig. 2b). The fitted values of $\beta$ are $-2.0 \pm 0.2 \, eV \cdot nm \cdot V^{-1}$ and $4.7 \pm 0.2 \, eV \cdot nm \cdot V^{-1}$ for $A_2^I$ and $A_2^{II}$, respectively.

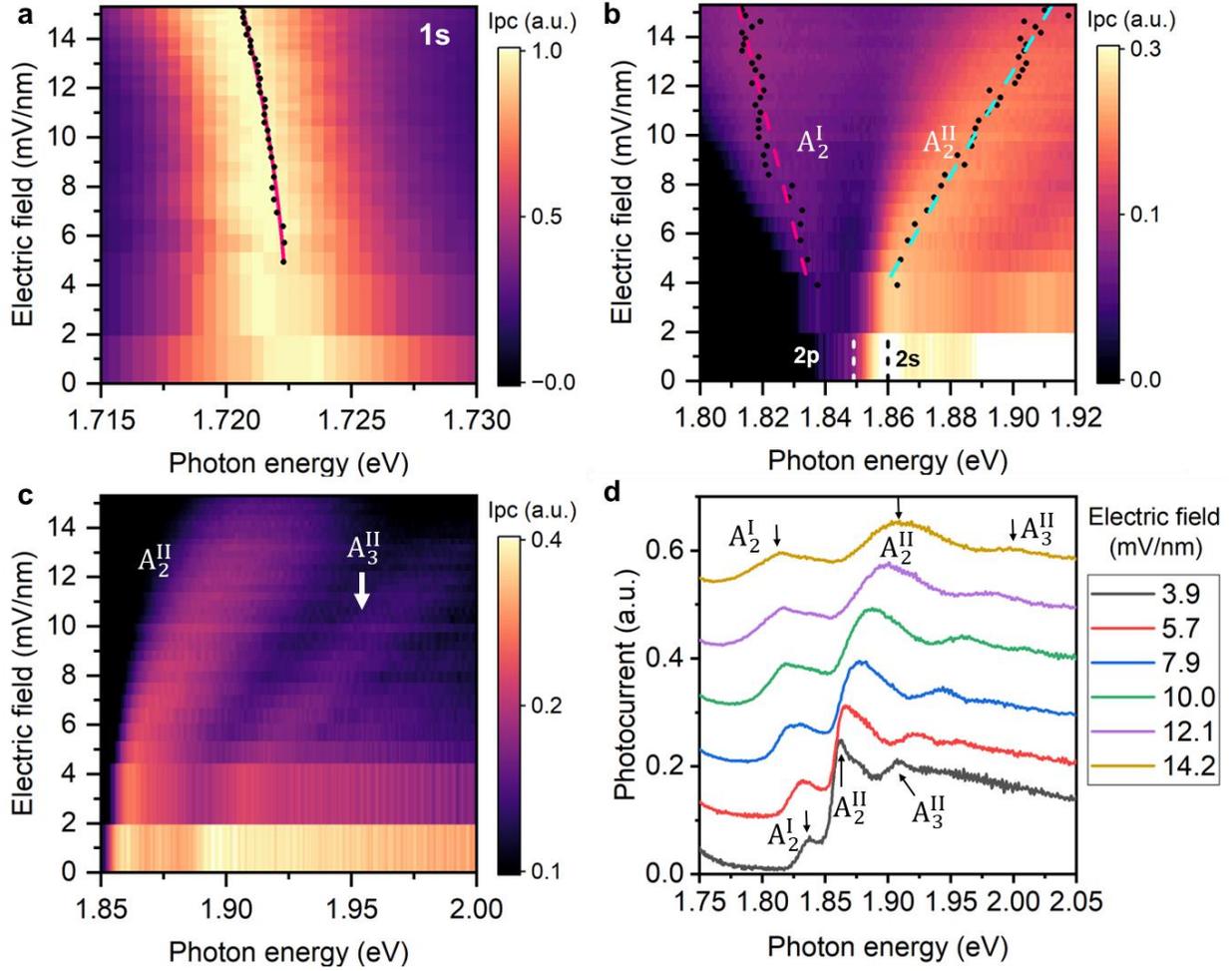

**Figure 2. Excitonic Stark effects in the WSe$_2$ p-n junction.** (a), (b) and (c) are the color plots of the photocurrent spectra as a function of the in-plane electric field for Rydberg states corresponding to n=1, 2, and 3, respectively. The black dots denote the extracted peak positions. The red line in (a) is the fitting of the 1s peak position to a quadratic function of the electric field $F$. The red and the cyan dashed lines in (b) are the fitting of the $A_2^I$ and $A_2^{II}$ peak positions to a linear function of the electric field, respectively. (d) shows the line cuts of the photocurrent spectra at six different electric field values. The temperature is 77 K for all the measurements shown in Fig. 2. The magnitude of the photocurrent in (a) – (d) is normalized by the magnitude of the 1s exciton.

We attribute the shifting and the splitting of the excitonic peaks to the excitonic Stark effects, which are a result of energy shift due to the external electric field modified exciton wavefunctions[9,10,26–30]. To quantitatively evaluate these effects, we calculate the energy of each excitonic state using the non-hydrogenic Keldysh potential[31,32] as the effective Coulomb interaction. Following the approach outlined in our previous work[11], we obtained the energies of the s, p, and d states for Rydberg excitons with n⩽3 (calculation details in methods and SI section 4), as shown in Table 1. We note that under zero electric field, only the s states are the bright states. Our result shows the 1s and 2s states possess a binding energy of 161.2 meV and 37.4 meV, respectively. The optically dark 2p state lies slightly below the 2s state and has a binding energy of 48.3 meV. It is worth noting that the 2p and 2s degeneracy is lifted in 2D due to the Keldysh potential, compared with the 2D hydrogen model[17]. It is also worth mentioning that the 2p lies at a lower energy than 2s, a result of the reduced screening for more spread wavefunction of 2p[3]. The same bandgap, together with the larger binding energy of 2p than 2s, makes the 2p resonance lower in energy, which is a unique feature of 2D excitons. The same argument can be extended to n=3 states, which we will discuss later.

|  | $n = 1$ | $n = 2$ | $n = 3$ |
|---|---|---|---|
| $Eb_{ns}$ (meV) | 161.2 | 37.4 | 16.1 |
| $Eb_{np}$ (meV) |  | 48.3 | 19.1 |
| $Eb_{nd}$ (meV) |  |  | 20.5 |

Table 1. Calculated binding energies of Rydberg excitons with s, p, and d orbitals at zero electric field.

We then quantitatively investigate the energies of Rydberg excitons in the presence of an in-plane electric field. The Hamiltonian describing the interaction between a 2D exciton and an in-plane electric field be written as:

$$H^e = -eFx \qquad (1),$$

where $e$ is the elementary charge, $F$ is the electric field, and $x$ is the coordinate in the direction of the electric field.

As the 1s exciton is well separated from the states with a higher principal number, its energy in the presence of an external in-plane electric field can be calculated by the second-order perturbation theory, which yields:

$$E_{1s} = E_{1s}^0 - 2\sum_n \frac{|\langle 1s|H^e|np\rangle|^2}{E_{np}^0 - E_{1s}^0} = E_{1s} - \alpha F^2 \qquad (2)$$

where $|np\rangle$ denotes the p-state exciton wavefunction with principal number n. $E_{np}^0$ and $E_{1s}^0$ denote the energy of np state and 1s state at zero electric field, respectively (details in SI section 5).

Based on Eqn. (1), the value of $\alpha$ is around 9.92 eV/(V/nm)$^2$, consistent with the reported values from previous theoretical $(9.8 - 10.4 \text{ eV/(V/nm)}^2)$[9,29,30] and experimental $(10 \pm 2 \ eV/(V/nm)^2)$[9] works, and also agrees well with the value of $8.0 \pm 0.2$ eV/(V/nm)$^2$ extracted from our experiment.

For states with n>1, their energy separations are comparable to the large Stark energy shift. As a result, the electric field is strong enough to induce a hybridization between multiple excitonic Rydberg states with distinct n and l numbers. Using the exciton wavefunctions at the zero electric field as bases, the interaction term is given by $\langle n'l'|H^e|nl\rangle$. The nature of $A_2^I$ and $A_2^{II}$ can be qualitatively understood with the hybridization between the 2p state and the 2s state, while the details could be revealed more clearly from the photocurrent spectra at a lower temperature, as discussed below.

To resolve the fine features resulting from the Stark splitting corresponding to n=3, we measured the electric field dependence of the photocurrent spectra at the temperature of 6 K, as shown in Figs. 3a, b. At the higher-energy side of $A_2^{II}$, we observed three high-energy peaks, denoted as $A_3^I$, $A_3^{II}$ and $A_4$, which shift linearly under an electric field, with their slopes being $0.5 \pm 0.2\ eV \cdot nm \cdot V^{-1}$, $7.7 \pm 0.2\ eV \cdot nm \cdot V^{-1}$ and $12.4 \pm 0.7\ eV \cdot nm \cdot V^{-1}$, respectively. The low-energy branch $A_3^I$ emerges at around 17 meV below $A_3^{II}$ at a relatively low electric field of 2 mV/nm, while the high-energy branch $A_4$ is resolvable at an electric field above 4 mV/nm.

We now develop a quantitative understanding of the orbital hybridization effect. Using the original $2p_x$, 2s, $3d_{x^2-y^2}$, $3p_x$, and 3s orbitals as the bases, we solve the eigenvalues and eigenfunctions of the interaction Hamiltonian, which lead to the resonance energies of the hybridized excitons as a function of the electrical field shown as the lines in Fig. 3c. As the electric field introduces the mixing of different orbital components, the original $2p_x$, 2s, $3d_{x^2-y^2}$, $3p_x$, and 3s orbitals, labeled as 2p, 2s, 3d, 3p, and 3s at zero electric field (Fig. 3c), will mix and generate five dispersive curves as C1-C5 in Fig. 3c. At the high electric field limit, we also label the major component of C1, C2, and C3, with the calculated components of different orbital bases shown in SI section 6. To compare the calculation with our experiment data, we show the peak positions extracted from Fig. 3a (solid dots) along with C1-C5 in Fig. 3c. It can be found that our theoretical model can explain most features on the photocurrent spectra, including the opposite shift between $A_2^I$ and $A_2^{II}$ and the increased separation between $A_3^I$ and $A_3^{II}$ at the increased electric field. Below an electric field of 2 mV/nm, the angular quantum number l is approximately a good quantum number for these excitonic states. As the electric field tunes two states close in energies, the interaction results in the transferring of their main orbital components. Our results show that the field dependence of $A_2^I$ is consistent with the curve C2. At the electric field as high as 8 mV/nm, the C2 is still visible in the photocurrent spectra due to the finite oscillator strength inherited from the original 2s and 3s orbitals, despite the main component being 2p. $A_2^{II}$ is mainly from the 2s orbital, corresponding to curve C2 for $F < 2\ mV/nm$ and C3 for $F > 2\ mV/nm$. Here, we did not observe the level avoiding between C2 and C3, likely due to the experimental uncertainty induced by the finite peak width. We note that, theoretically, there should be another dispersive branch, curve C1, below $A_2^{II}$. However, it is not observed in the photocurrent spectra, possibly because of the strong dissociation effect of the 3d orbital, which is the main component of C3 at the high electric field, despite finite oscillator strength transferred from 2s (see details in SI section 6). Also, the 3d nature might broaden the associated resonance significantly so that the finite oscillator strength distributed over a broad range of energy cannot be resolved from the already broad $A_2^I$ peaks as shown in Fig. 3b. The electric field dependence of $A_3^I$ and $A_3^{II}$ matches the dispersion of curve C4 and curve C5. The visibility of the $A_3^I$ is a result of the hybridization of the original 3s, 3p, and even 2s states, which transfer some oscillator strength from 3s and 2s to C4 at finite electric field and make the original dark 3p state (C4 at zero electric field) bright. The energy of $A_4$ is obviously higher than the predicted energies of states with n=3, suggesting that it corresponds to an excitonic state with a higher n number. The $A_4$ peak is significantly broader and weaker in strength than other resonances, and we refrain from quantitative understanding in this work. It is worth noting that at the field ranging from 6 mV/nm to 9 mV/nm, Rydberg exciton with principal number n=4 is likely to be dissociated. The nature of $A_4$ is intriguing to be explored in the future.

In summary, we have employed photocurrent spectra to probe the Rydberg exciton resonances from a monolayer WSe$_2$ p-n junction. We have observed significant Stark effects on Rydberg excitons with principal numbers up to n=3. The electric field introduces significant mixing of different orbitals and brightens the otherwise dark states such as 3p and 3d. The electric field, therefore, induces a large energy shift observable in the photocurrent spectra, more than one order of magnitude larger than of 1s exciton. Our works thus pave the way for utilizing Rydberg excitons in the application of quantum sensing.

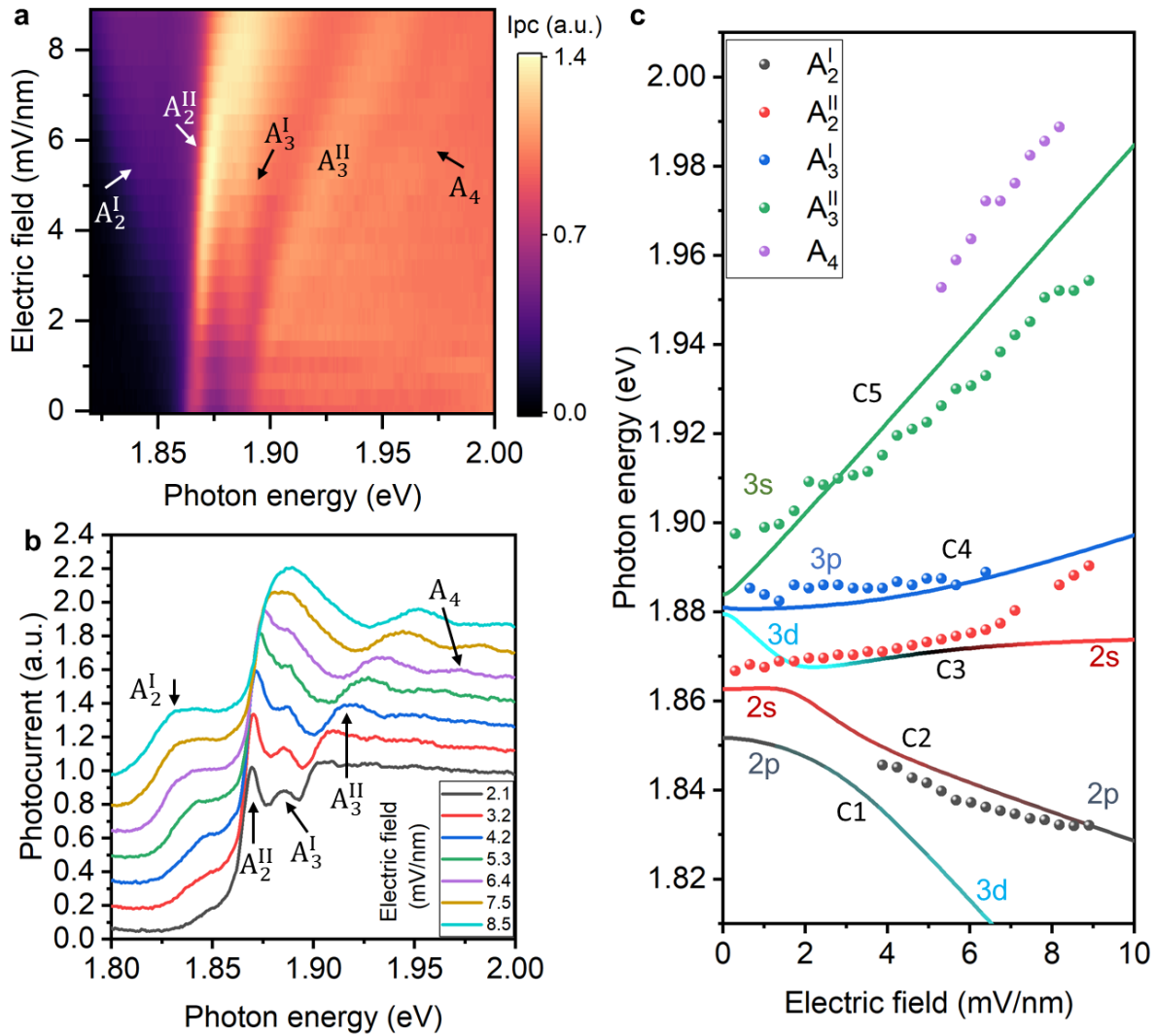

Figure 3. Excitonic Stark effects of Rydberg excitons with n = 3. (a) shows the colormap of the photocurrent spectra as a function of the electric field near the 3s state of monolayer WSe$_2$ taken at 6 K. (b) shows the linecuts at seven electric field values from (a). The dots in (c) are the peak positions extracted from (a). The dashed lines are the theoretical exciton energies calculated by considering the hybridization between multiple exciton orbitals.

## Methods

### Device fabrication

The patterns of the split-gate electrodes were defined using the standard electron-beam lithography technique using PMMA as resist. 3 nm Ti and 22 nm Au were deposited using electron-beam evaporation. After metal deposition, a lift-off process was performed in acetone and isopropanol. The 2D material flakes were exfoliated onto Polydimethylsiloxane (PDMS) films. The bottom h-BN, monolayer $WSe_2$ (from HQ graphene), few-layer graphene, and top h-BN flakes were transferred on to the electrode layer by layer using a dry-transfer technique[33] reported previously. The optical image of the device can be found in Supplementary Section 1.

### Photocurrent measurement

To perform photocurrent spectroscopy, the sample was mounted into a cryostat with an optical window and electrical feedthrough. Broadband white light with a repetition rate of 10 MHz and a pulse width of 100 ps was generated using a supercontinuum laser source (YSL Photonics) and focused into a monochromator (Princeton Instruments). A particular wavelength was selected using a diffraction grating. The monochromatic light at the exit of the monochromator was collimated using a lens and then focused into a spot with a diameter of around 4 μm on the sample using an objective (50X, NA = 0.55). A mechanical chopper was used to apply AC modulation to the incident laser. The gate voltages on the sample were applied using Keithley 2400 multimeters. The photocurrent from the sample was collected using a preamplifier (Stanford Research Systems), a lock-in amplifier (Stanford Research Systems) and a data acquisition card (National Instruments).

### Electric field calculation

The in-plane electric field as a function of gate voltages at 77 K was calculated using an analytical method developed by previous works[9,22,23]. The monolayer $WSe_2$ was first treated as a metal plate to obtain the charge density distribution. The electric field is then calculated by solving the Laplace equation in the 2D channel using the boundary conditions defined by the charge distribution. The details of this calculation can be found in SI section 2. Due to the increased contact resistance at 6 K, the electric field in Fig. 3 was calibrated by interpolating the $A_2^I - A_2^{II}$ splitting obtained from Fig. 2 (See Supplementary Section 3 for details).

### Calculation of exciton energies

The energy of each excitonic Rydberg state at zero electric field was calculated using the same method described by our previous work[11]. We numerically solved the two-body Schrodinger equation in two-dimension using the nonhydrogenic Keldysh potential[31,32] as the effective Coulomb interaction. The eigenstates and eigenenergy were obtained numerically by expanding the wavefunction in terms of Bessel bases. The bandgap of $WSe_2$ ($E_g$) was set to be 1.90 eV. The surrounding dielectric constant $\varepsilon$, and the 2D polarizability $\chi_{2D}$ in the Keldysh potential are $4.5\varepsilon_0$ and 0.718 nm, respectively. The reduced mass of the exciton $m_r$ is $0.2m_0$, where $m_0$ is the electron mass. The energy of the 1s state under an in-plane electric field was calculated using the second-order perturbation theory. The energies of the Rydberg states with $n = 2$ and 3 were calculated by considering the electric field-induced hybridization of s, p, and d orbitals, using the $2p_x$, 2s, $3d_{x^2-y^2}$, $3p_x$ and 3s orbitals as the bases. The calculation details can be found in SI sections 4 and 5.

## Acknowledgments

We thank Prof. Yong-Tao Cui for the helpful discussions. This work is mainly supported by NSF Grant ECCS- 2139692. S.-F. S. also acknowledges support from NSF Grants DMR-1945420, DMR-2104902, and NYSTAR through Focus Center-NY–RPI Contract C180117. The device fabrication was supported by the Micro and Nanofabrication Clean Room (MNCR) at Rensselaer Polytechnic Institute (RPI). Y.-M. L. acknowledges support from Grant No. 2022YFA1204700 from the Ministry of Science and Technology of the People's Republic of China. K. W. and T. T. acknowledge support from JSPS KAKENHI (Grant Numbers 19H05790, 20H00354, and 21H05233). The optical spectroscopy measurements were supported by the AFOSR DURIP awards through Grants FA9550-20-1-0179 and FA9550-23-1-0084. C. Z. acknowledges support from NSF (PHY-2409943, OMR-2228725, ECCS-2411394) and AFOSR (FA9550-20-1-0220).

## Author contributions

S.-F. S. conceived the project. Z. L. fabricated devices and performed measurements. Y.-M. L. and C. Z. performed the calculations. T. T. and K. W. grew the BN crystals. S.-F. S, Z. L, L. Y., and L. M. analyzed the data. S.-F. S. supervised the project. S.-F. S. and Z. L. wrote the manuscript with input from all authors.

## Competing interests

The authors declare no competing interest.